\documentclass {elsart}

\usepackage{graphicx,natbib,amssymb}

\usepackage[centertags]{amsmath}

\numberwithin{equation}{section}

\begin{document}

\begin{frontmatter}

\title{A step beyond Tsallis and R\'{e}nyi entropies}

\author{Marco Masi\corauthref{cor}} \corauth[cor]{Corresponding author.}
\ead{marco.masi@spiro.fisica.unipd.it, marco\!\_masi2@tin.it}
\address{Dipartimento di Fisica G. Galilei, Padova, Italy.}

\begin{abstract}
Tsallis and R\'{e}nyi entropy measures are two possible different generalizations of the
Boltzmann-Gibbs entropy (or Shannon's information) but are not generalizations of each others. It
is however the Sharma-Mittal measure, which was already defined in 1975 (B.D. Sharma, D.P. Mittal,
J.Math.Sci \textbf{10}, 28) and which received attention only recently as an application in
statistical mechanics (T.D. Frank \& A. Daffertshofer, Physica A \textbf{285}, 351 \& T.D. Frank,
A.R. Plastino, Eur. Phys. J., B \textbf{30}, 543-549) that provides one possible unification. We
will show how this generalization that unifies R\'{e}nyi and Tsallis entropy in a coherent picture
naturally comes into being if the q-formalism of generalized logarithm and exponential functions
is used, how together with Sharma-Mittal's measure another possible extension emerges which
however does not obey a pseudo-additive law and lacks of other properties relevant for a
generalized thermostatistics, and how the relation between all these information measures is best
understood when described in terms of a particular logarithmic Kolmogorov-Nagumo average.
\end{abstract}


\end{frontmatter}

\newpage

\section{Introduction}

To gain a unified understanding of the different entropy measures and how they relate to each
others in the frame of a generalized picture, it is first necessary to recall what characterizes
"classical" entropies and emphasize some aspects which are important for the present paper.

\subsection{The Boltzmann-Gibbs entropy and Shannon's information measure}

As it is well known, given a probability distribution $P= \{p_{i}\}$, $(i=1,...,N)$, with $p_{i}$
representing the probability of the system to be in the i-th microstate, the Boltzmann-Gibbs (BG)
entropy reads $$S_{BG}(P)= - k \, \sum^{N}_{i=1} p_{i} \log p_{i} \, ,$$ where $k$ is the
Boltzmann constant and $N$ the total number of possible configurations. If all states are
equi-probable it leads to the famous Boltzmann principle $S = k \, \log W$ (N=W). BG entropy is
equivalent to Shannon's expression if we set $k=1$ (as we will do from now on) and use the
immaterial base $b$ for the logarithm function $$S_{S}(P) = - \sum^{N}_{i=1} p_{i} \log_{b} p_{i}
\, .$$ It is common to use the natural base for the BG entropy, while base 2 has the advantage to
deliver information quantities in bits.

What characterizes BG and Shannon's measure is additivity of information. Given two systems,
described by two independent probability distributions $A$ and $B$ (i.e. $P(A \cap B) = P(A)
P(B)$), using an additive information measure means that $$S_{S}(A \cap B) = S_{S}(A) + S_{S}(B|A)
\, , $$ with $$S_{S}(B|A) = \sum_{i} p_{i}(A) \, S_{S}(B|A=A_{i}) \, ,$$ being the conditional
entropy. In this case we are talking about \textit{extensive systems}, i.e. systems where the
entropy is given by the sum of all the entropies of their parts, as it is customary to do in
standard statistical mechanics. The unique function which assures additivity is the logarithm.
Also in the axiomatic derivation of Shannon's entropy performed by A.I. Khinchin \cite{Khinchin},
it is the additive property which leads to the appearance of the logarithm function. This is the
real reason that stands behind the ubiquitous presence of the logarithm function in information
theory, and we can confidently say that every modification to it reflects a deviation from the
additive law.

We will from now on use the natural base. Shannon's entropy can be written in the form of a
"linear" (the arithmetic) mean as \begin{equation} S_{S}(P) = \left< I_{i}\right>_{\mathrm{lin}} =
\left< \log \left( \frac{1}{p_{i}} \right) \right>_{\!\mathrm{lin}} \,, \label{S2}
\end{equation} where we will call the quantity $$I_{i}=\log \left( \frac{1}{p_{i}}
\right) \, ,$$ the \textit{elementary information gain} associated to an event of probability
$p_{i}$ (in information theory it is sometimes called the \textit{code length}). The quantity
$\frac{1}{p_{i}}$ is also called the \textit{surprise} (less probable events are considered more
"surprising" than more probable ones), and we will see that it is this quantity which is really
measured in one way or another, not $-\log p_{i}$.

\subsection{Tsallis' entropy}

Additivity is however not always preserved, especially in nonlinear complex systems, e.g. when we
have to deal with long range forces, as it is in the case of the dynamic evolution of star
clusters or in systems with long range microscopic memory, in fractal- or multifractal-like and
self-organized critical systems, etc. We are dealing in this case with \textit{non-extensive
systems}; a case which received much attention in the last decade.\cite{Tsallis2}

A generalization of the BG entropy to \textit{non-extensive systems} is known as Tsallis entropy
\cite{Tsallis}. C. Tsallis noted that if non-extensivity enters into the play things are described
better by power law distributions, $p_{i}^{\,q}$, so called \textit{q-probabilities}, i.e. by
scaled probabilities where $q$ is a real parameter. This introduces the formal possibility not to
set rare and common events on the same footing, as in BG or Shannon statistics, but it enhances or
depresses them according to the parameter chosen (in complex systems rare events can have dramatic
effects on the overall evolution).

With the introduction of the normalized \textit{q-probabilities} it became customary to define so
called \textit{escort-} or \textit{zooming-distribution} $$\pi_{i}(P, q) =
\frac{p_{i}^{q}}{\sum_{i=1}^{N} p_{i}^{q}} \, ; \hspace{10mm} q>0 , \,\,\, q \in \Re .$$

In this frame Tsallis postulated his now famous generalization of Shannon's entropy to
non-extensivity  \cite{Tsallis}:

\vspace{-2mm}

\begin{equation}
S_{T}(P,q) = \frac{\sum_{i=1}^{N}p_{i}^{q}-1}{1-q} \, = \, \frac{1}{q-1} \sum_{i=1}^{N}p_{i} \,
(1- p_{i}^{q-1}) \, .\label{tsallis}
\end{equation}

For $q\rightarrow1$, Shannon's measure is recovered, i.e.: $S_{T}(P,1) = S_{S}(P) \, .$

Tsallis entropy extends to a \textit{pseudo-additive} law \begin{equation}S_{T}(A \cap B) =
S_{T}(A) + S_{T}(B|A) + (1-q)S_{T}(A) S_{T}(B|A) \, ,\label{padditive}\end{equation} with
$$S_{T}(B|A) = \sum_{i}\pi_{i}(A) \, S_{T}(B|A=A_{i}) \, .$$

Let us introduce the \textit{generalized q-logarithm function}
\begin{equation} \log_{q} x = \frac{x^{1-q}-1}{1-q} \, , \label{q-log} \end{equation} which, for $q=1$,
becomes again the common natural logarithm. Its inverse is the \textit{generalized q-exponential
function} \begin{equation} e_{q}^{x} = [1+(1-q)x]^{\frac{1}{1-q}} \, ,\label{q-exp}
\end{equation} which becomes the exponential function for $q=1$. The importance of the q-logarithm is
that it satisfies a pseudo-additive law \begin{equation}\log_{q} xy = \log_{q} x + \log_{q} y +
(1-q) (\log_{q} x) (\log_{q} y) \, .\label{pseadd} \end{equation} Then Tsallis entropy
\ref{tsallis} can be written as the \textit{q-deformed Shannon entropy} \begin{equation}
S_{T}(P,q) = - \sum_{i=1}^{N}p_{i}^{q} \log_{q}p_{i} = \sum_{i=1}^{N} p_{i}
\log_{q}\left(\frac{1}{p_{i}}\right) = \left< \log_{q}\left(\frac{1}{p_{i}}\right)
\right>_{\!\mathrm{lin}}= \left< I_{i} \right>_{\!\mathrm{lin}} \, , \label{q-shannon}
\end{equation} with the last term resulting as the q-extension of \ref{S2}. This reflects the
non-extensive character of the system on the elementary information gains.

Note also that the classical power laws and the additivity rules for the logarithm and exponential
do no longer hold in this generalized context. Except for $q=1$, in general $\log_{q}x^{\alpha}
\neq \alpha \log_{q}x$, which explains why we keep writing throughout this paper Shannon's
elementary information gain as $ \left<\log \left(\frac{1}{p_{i}}\right)\right>_{\!\mathrm{lin}}$
instead of $-\left<\log p_{i}\right>_{\!\mathrm{\, lin}}$. Useful for our purposes will be the
equality \begin{equation}e_{q}^{x + y + (1-q)x y } = e_{q}^{x} \, e_{q}^{y} \, .\label{eqplus}
\end{equation} We will see how the q-deformed formalism fits naturally in the mathematical descriptions of
generalized entropy measures.

\subsection{R\'{e}nyi's entropy}

Either in the case of BG as for Tsallis entropy, in \ref{S2} and \ref{q-shannon}, an entropy
measure is the average S obtained over many \textit{elementary information gains} $I_{i} \equiv
I_{i}(\frac{1}{p_{i}}) = \log_{q}\left(\frac{1}{p_{i}}\right)$ associated to the i-th event of
probability $p_{i}$ (if the system is extensive, q=1).

Another possible generalization exists and has become commonplace throughout the literature,
namely R\'{e}nyi's measure \cite{Renyi}. A. R\'{e}nyi maintained a still additive measure, as in
BG entropy, but considered that another form of averaging is possible. His starting point was the
generalized notion of average of A.N. Kolmogorov and M. Nagumo (\cite{Kolmogorov2},
\cite{Nagumo}), who independently showed that, in the frame of the Kolmogorov axioms of
probability theory, the definition of the average must be extended to the
\textit{quasi-arithmetic} or \textit{quasi-linear mean} defined as \begin{equation} S = f^{-1}
\left( \sum_{i=1}^{N} p_{i} \, f(I_{i}) \right) \, , \label{infomean} \end{equation} where $f$ is
a strictly monotone continuous and invertible function, the so called \textit{Kolmogorov-Nagumo
function} (KN function). On his side, R\'{e}nyi showed that if we restrict to additive measures
then only two possible KN functions exist. The first one is the common arithmetic mean and is
associated with the KN function $f(x) = x$, and the second is the \textit{exponential mean} with
\begin{equation}f(x) = c_{1} \, b^{(1-q)x} + c_{2} \, , \label{quasilin} \end{equation} where $q$ is a
real parameter, and $c_{1}$ and $c_{2}$ are two arbitrary constants.

The exponential mean leads to \textit{R\'{e}nyi's information measure} or \textit{R\'{e}nyi's
entropy} \begin{equation} S_{R}(P,q) = \frac{1}{1-q} \log_{b} \sum_{i=1}^{N} p_{i}^{q} \, ,
\label{renyi} \end{equation} with $b$ the logarithm base (we will from now on assume the natural
base, $b=e$, for R\'{e}nyi's entropy either). For $q\rightarrow1$ R\'{e}nyi's measure becomes
Shannon's entropy.

It should be noted how P. Jizba and T.Arimitsu \cite{jarenyi} showed that R\'{e}nyi's measure can
be obtained also extending the Shannon-Khinchin axioms to a quasi-linear conditional information
\begin{equation}S_{R}(B|A)=f^{-1}\left( \sum_{i} \pi_{i}(A) f \left(S_{R}(B|A_{i})\right)
\right)\, , \label{irba} \end{equation} with $f$ as given in \ref{quasilin}.

Therefore Shannon's information measure is an averaged information in the ordinary sense, while
R\'{e}nyi's measure represents an exponential mean over the same elementary information gains
$\log \left(\frac{1}{p_{i}}\right)$.

\section{The Sharma-Mittal and \textit{Supra-extensive} entropy}

\subsection{Generalizing with Kolmogorov-Nagumo means}

It is important to understand that Tsallis and R\'{e}nyi entropies are two different
generalizations along two different paths. Tsallis generalized to non-extensive systems, while
R\'{e}nyi to quasi-linear means. But we can search for an entropy which generalizes to
non-extensive sets and non-linear means containing Tsallis and R\'{e}nyi measures as limiting
cases.

Let us unify the picture of all the entropies considered here through KN averages (as J.Naudts and
M.Czachor did \cite{Naudts}, tough by a slightly different approach).

It is immediate to see from \ref{q-shannon} and \ref{infomean} how for Tsallis's measure it is the
KN function \begin{equation}f(x)= x \label{tsallisfun}\end{equation} which averages over the
elementary information gain $$I_{i}=\log_{q} \left(\frac{1}{p_{i}}\right) \, .$$ This led us to
write it as $$S_{T}(P,q) = \left< \log_{q}\left(\frac{1}{p_{i}}\right) \right>_{\!\mathrm{lin}} \,
.$$

While, for R\'{e}nyi's measure, choose in \ref{quasilin}, $c_{1}=\frac{1}{1-q}=-c_{2}$ (remember
\ref{q-log}), then the KN function takes the form
\begin{equation}f(x) = log_{q} \, e^{x} \, ,\label{renyifun} \end{equation} which, applied on
$$I_{i}=\log \left(\frac{1}{p_{i}}\right) \, ,$$ in \ref{infomean} ($f^{-1}(x) = \log e_{q}^{x}$)
leads us to rewrite \ref{renyi} as $$S_{R}(P,q) = \left< \log \left( \frac{1}{p_{i}} \right)
\right>_{\!\!\mathrm{exp}} \, ,$$ where, of course, $\left< \, . \, \right>_{\mathrm{exp}} \equiv
\left< \, I_{i} \, \right>_{\mathrm{exp}}$ stands for the exponential mean defined by the KN
function \ref{renyifun} over the elementary information $I_{i}$.

But, what Tsallis and R\'{e}nyi measures have in common is that in both cases
\begin{equation}f(I_{i})=\log_{q} \left( \frac{1}{p_{i}} \right) \, . \label{f(i)}\end{equation}

Then, for a further generalization, the simplest step beyond them would be that to generalize
\ref{tsallisfun} and \ref{renyifun} with
\begin{equation}f(x) = log_{q} \, e_{r}^{x} \, \label{naudtsfun} \end{equation} and set $$ I_{i}=\log_{s}
\left(\frac{1}{p_{i}}\right) \, \, ,$$ where $r,s$ are new parameters on the generalized
exponential and logarithm functions. Maintaining constraint \ref{f(i)} implies $s=r$. Then
calculating \ref{infomean} ($f^{-1}(x) = log_{r} \, e_{q}^{x})$, one obtains the
\textit{Sharma-Mittal information measure} \cite{Sharma}\begin{equation} S_{SM}(P,\{q,r\}) =
\log_{r}e_{q}^{\sum_{i}p_{i}\log_{q}\left(\frac{1}{p_{i}}\right)}= \label{naudtslog}
\end{equation}
\vspace{2mm}
$$=\left<\log_{r}\left(\frac{1}{p_{i}}\right) \right>_{\!\!\mathrm{q-exp}}=
$$ \vspace{2mm}
$$= \frac{1}{1-r} \left[ \left( \sum_{i} p_{i}^{\,q} \right) ^{\frac{1-r}{1-q}} - 1 \right]\, ,$$

where $\left<\,\cdot\ \right>_{q-exp}$ stands for an average defined by the KN function
\ref{naudtsfun} and that we will call the \textit{quasi-exponential mean}.

We can see that for $r \rightarrow 1$ R\'{e}nyi's measure, and for $r \rightarrow q$ Tsallis
measure, are recovered as limiting cases.

We will show in the next section that for two statistical independent systems $A$ and $B$ it is
easy to check that $$S_{SM}(A \cap B) = S_{SM}(A) + S_{SM}(B|A) + (1-r)S_{SM}(A) S_{SM}(B|A) \, ,
$$i.e. a pseudo-additive law holds as in the case of Tsallis entropy.

Therefore Sharma-Mittal's measure generalizes R\'{e}nyi's extensive entropy to non-extensivity,
characterized by the r-logarithm. It is the parameter $r$ which determines the degree of
non-extensivity, while $q$ is the deformation parameter of the probability distribution (however,
when $r \rightarrow q$ the two parameters become intertwined and in Tsallis entropy it is q which
measures non-extensivity).

On information theoretic grounds, B.D. Sharma and D.P. Mittal \cite{Sharma}, advanced already in
1975 this non-additive measure which shows to have a non-extensive character either. But it wasn't
until recently (\cite{Daffertshofer}, \cite{Frank}, and without mentioning it explicitly
\cite{Naudts}) that Sharma-Mittal's measure has been investigated in statistical mechanics.

\subsection{Generalizing with q-logarithms and q-exponentials}

At this point let us see how by using the q-deformed logarithm and exponential formalism, one
could express in a much more compact form the same generalization path.

First of all recall a well known relationship which exists between Tsallis and R\'{e}nyi
entropies, namely \begin{equation} S_{R}(P,q) =\frac{1}{1-q} \log \left[ 1+ (1-q) \, S_{T}(P,q)
\right] \, . \label{irit} \end{equation}

Here we can efficiently exploit the generalized logarithm and exponential functions \ref{q-log}
and \ref{q-exp}, rewriting \ref{irit} in the more compact form \begin{equation}S_{R}(P,q) = \log
e_{q}^{S_{T}(P,\,q)} \, ,\label{ireqt} \end{equation} from which follows immediately
\begin{equation}S_{T}(P,q) = \log_{q}e^{S_{R}(P,\,q)} \, . \label{itqir}
\end{equation}

Looking at the structure of \ref{ireqt} and \ref{itqir} we can ask if, given another parameter
$r$, the following \begin{equation}S_{SM}(P,\{q,r\})=\log_{r} e_{q}^{S_{T}(P,\,q)} = \frac{1}{1-r}
\left[ \left( \sum_{i} p_{i}^{\,q} \right) ^{\frac{1-r}{1-q}} - 1 \right] \, , \label{eqit}
\end{equation} and \begin{equation}\hspace{9mm} S_{SE}(P,\{q,r\})= \log_{q}e_{r}^{S_{R}(P,\,q)} =
\frac{\left[ 1 + \frac{(1-r)}{(1-q)} \, \log \sum_{i}p_{i}^{q}\right]^{\frac{1-q}{1-r}}-1}{1-q}\,
,\label{qrir}\end{equation}

might then be other possible generalizations? \ref{eqit} can be recognized immediately as
Sharma-Mittal's measure \ref{naudtslog} and can be already accepted as an extension.

\ref{qrir} instead needs a closer look. For $r \rightarrow q$ it obviously boils down to
R\'{e}nyi's entropy. For $r \rightarrow 1$ we obtain Tsallis' measure again. So, from a formal
point of view it can be regarded as another generalization too. It is however not entirely clear
what kind of statistics it expresses. Its particular status might be best evidenced expressing all
the measures in terms of (logarithmic averaged) surprise quantities.

Indeed, notice that we can rewrite the quantity
\begin{equation}\left(\sum_{i}p_{i}^{q}\right)^{\frac{1}{1-q}} \! \! \!= \left(\sum_{i}p_{i} \left(
\frac{1}{p_{i}} \right) ^{1-q}\right)^{\frac{1}{1-q}} \! \! \! = \left<
\left(\frac{1}{p_{i}}\right)^{1-q}\right>^{\frac{1}{1-q}}_{\!\mathrm{lin}} \! \! =
e_{q}^{\left<log_{q}\left(\frac{1}{p_{i}}\right)\right>_{\!\mathrm{lin}}} =
\left<\frac{1}{p_{i}}\right>_{\!\!\log_{q}} \, ,\label{eii}\end{equation} where we used the
logarithmic mean $\left< \cdot \, \right>_{log_{q}}$ defined by the KN function $f(x)=log_{q}x$.
Then, from \ref{q-shannon} and \ref{renyi}, and using \ref{eii}, equations \ref{ireqt} to
\ref{qrir} can be rewritten as \begin{equation} \hspace{-0mm} S_{R}(P,q) = \log
\left<\frac{1}{p_{i}}\right>_{\!\!\log_{q}} \, ;\label{nonrelrenyi}
\end{equation}

\begin{equation}\hspace{-0mm} S_{T}(P,q)  = \log_{q}
\left<\frac{1}{p_{i}}\right>_{\!\!\log_{q}} \, ;\label{nonreltsallis}\end{equation}

\begin{equation}\hspace{-0mm} S_{SM}(P,\{q,r\}) = \log_{r}
\left<\frac{1}{p_{i}}\right>_{\!\!\log_{q}}  \,
;\label{nonrelnaudts}\end{equation}

\begin{equation}\hspace{-0mm} S_{SE}(P,\{q,r\}) = \log_{q}
e_{r}^{\,\log \left<\frac{1}{p_{i}}\right>_{\!\log_{q}}} \, . \label{nonrelmasi}\end{equation}

With the q-deformed logarithm and exponential formalism we could easily see the generalization
path to follow and write all the measures into a more compact form (\ref{ireqt} to \ref{qrir}).
Moreover this makes it easier to recognize the behavior of the limits than in their explicit form
(the r.h.s. of \ref{eqit} and \ref{qrir}). With no or only few passages it is immediate to see how
\ref{eqit} reduces to Tsallis entropy for $r \rightarrow q$, and  for $r \rightarrow 1$ it reduces
to R\'{e}nyi's entropy (without any need to apply Hopital rule, first order approximations or
whatever, insert \ref{tsallis} in \ref{q-exp}).

The limit for $q\rightarrow 1$ for Sharma-Mittal measure is
$$\lim_{q\rightarrow1}S_{SM}= \lim_{q\rightarrow1} \log_{r} e_{q}^{S_{T}}=\log_{r} e^{S_{S}} =
\log_{r} \left<\frac{1}{p_{i}}\right>_{\log}  =$$ $$\hspace{40mm} = \frac{e^{-(1-r) \sum_{i} p_{i}
\log p_{i}}-1}{1-r} \, ,$$ which Frank and Daffertshofer \cite{Daffertshofer} used to call the
\textit{gaussian entropy}.

By the way $$\lim_{q\rightarrow1}S_{SE}= \lim_{q\rightarrow1}\log_{q} e_{r}^{S_{R}} = \log
e_{r}^{S_{S}} = \log e_{r}^{\log \left<\frac{1}{p_{i}}\right>_{\!\!\log}} =$$ $$= \frac{1}{1-r} \,
\log \left( 1-(1-r)\sum_{i}p_{i} \log p_{i} \right) \, .$$

But the point is that when compared with \ref{nonrelrenyi}, \ref{nonreltsallis},
\ref{nonrelnaudts}, measure \ref{nonrelmasi} seems to stand apart and does not correspond to some
quasi-linear mean in the style of \ref{infomean}.

\subsection{Comparing the supra-extensive entropy with Sharma-Mittal's entropy}

Let us then focus shortly on the separate nature of \ref{nonrelmasi} (or \ref{qrir}) and some of
its properties.

First of all note that it can be shown how for two statistical independent systems $A$ and $B$,
similarly to Tsallis' entropy, the Sharma-Mittal's entropy obeys a pseudo-additive law and can be
decomposed as in \ref{padditive}. It is almost immediate to see this by employing the generalized
exponential formalism. Thanks to \ref{padditive}, \ref{pseadd}, \ref{eqplus}, starting from the
middle term of \ref{eqit} we can write $$S_{SM}(A \cap B) = \log_{r}e_{q}^{S_{T}(A \cap B)} =$$
$$= \log_{r}e_{q}\left[S_{T}(A) + S_{T}(B|A) + (1-q) \, S_{T}(A)S_{T}(B|A)\right] =
\log_{r} \left(e_{q}^{S_{T}(A)}e_{q}^{S_{T}(B|A)}\right) =$$ $$=\log_{r}e_{q}^{S_{T}(A)} +
\log_{r}e_{q}^{S_{T}(B|A)} + (1-r)\log_{r}e_{q}^{S_{T}(A)}\log_{r}e_{q}^{S_{T}(B|A)} = $$
\begin{equation}=S_{SM}(A) + S_{SM}(B|A) + (1-r)S_{SM}(A) S_{SM}(B|A) \, .\label{psadd} \end{equation}

Proceeding in the same manner with \ref{qrir} leads however not to the same decomposition. Because
of R\'{e}nyi's measure additive character one can't go further than $$S_{SE}(A \cap B) =
\log_{q}e_{r}^{S_{R}(A \cap B)} = \log_{q} e_{r}\left[S_{R}(A)+S_{R}(B|A)\right] \,$$ with
$S_{R}(B|A)$ as given in \ref{irba}.

Entropy \ref{nonrelmasi} therefore obeys a new form of non-extensivity, we call
\textit{supra-extensivity}.

There are also other aspects which should be mentioned. Let us briefly recall the notions of
concavity and stability applied to entropy measures.

Given two probability distributions $P=\{p_{1},...,p_{N}\}$ and $P'=\{p'_{1},...,p'_{N}\}$ and
defining an \textit{intermediate distribution} $P''=\{p''_{1},...,p''_{N}\}$ with $$p''_{i} \equiv
\mu p_{i} + (1-\mu)\,p'_{i} \, ; \hspace{5mm} \forall \mu \in [0,1] \, ,$$ $S(P)$ is said to be a
\textit{concave entropic functional} if and only if $$S(P'') \geq \mu S(P) + (1-\mu) S(P') \, .$$
Otherwise, $S(P)$ is said to be \textit{convex}. Concavity implies thermodynamic stability (e.g.
thermal equilibrium between two initial temperatures in BG statistical mechanics).

Recall also the notion of \textit{stability} (or \textit{experimental robustness}, as Tsallis
calls it, in order to avoid confusion with the previous form of thermodynamic stability) which
implies that for arbitrary small variations of the probabilities $p_{i}$ a statistical functional
remains finite. That is, given a \textit{deformation} $$||p-p'|| = \sum_{i} |p_{i}-p'_{i}| \, ,$$
such that $||p-p'|| < \delta_{\varepsilon}$, we obtain stability of $S(P)$ if $$\triangle = \left|
\frac{S(P)-S(P')}{S_{max}} \right| < \varepsilon \, ; \hspace{5mm} \forall \delta_{\varepsilon}>0,
\forall \epsilon >0 \, ,$$ with $S_{max}$ the maximum value $S$ can attain and for all microstates
$i=1,...,N$.

Lesche claims \cite{Lesche} that this is a necessary condition for an entropy measure to be a
physical quantity and showed that, while BG entropy is always stable, R\'{e}nyi's measure is
unstable for all $q \neq 1$. It is also known that BG entropy is always concave, while R\'{e}nyis
measure is concave only for $q \leq 1$, and can be either concave or convex for $q > 1$. More
recently, Abe \cite{Abe} showed that Tsallis entropy is concave and stable for all positive values
of $q$. It might also be worth mentioning that a physical entropy is not only expected to be
generically concave and Lesche-stable but should lead also to a finite entropy production per unit
time. BG and Tsallis entropies share all these properties. R\'{e}nyi entropy shares none.
\cite{Gell-Mann}

So, being an extension of it, it is clear that the properties of concavity and stability and
finite entropy production per unit time are generically violated also in Sharma Mittal's and the
supra-extensive entropy.

Finally, it should also be underlined how Frank and Plastino showed \cite{Frank} that the
Sharma-Mittal entropy is the only measure that allows for a pseudo-additive decomposition and at
the same time gives rise to a thermostatistics based on escort mean energy values $$U =
\frac{\sum_{i} p_{i}^{\,q} \varepsilon_{i}}{\sum_{i} p_{i}^{\,q}} \, ,$$ ($\varepsilon_{i}$ are
the energy levels) admitting of a generalized partition function $\widetilde{Z}$ defined by
$\log_{r}\widetilde{Z}_{SM}:= \log_{r}Z_{SM} - \beta U$ with \begin{equation}Z_{SM}=\left(
\sum_{i} p_{i}^{\,q}\right)^{\frac{1}{1-q}} = \left<\frac{1}{p_{i}}\right>_{\log_{q}} \, ,
\label{partz}\end{equation} ($Z_{SM}$ is the partition function which takes $U$ while
$\widetilde{Z}_{SM}$ takes zero as the energy reference, and where $\beta$ is an inverse
temperature measure), that leads to the usual expressions for the free energy
$$F=U - TS_{SM} = -\frac{1}{\beta} \log_{r}\widetilde{Z}_{SM}$$ and the mean energy $$U = - \frac{\partial}{\partial
\beta} \log_{r}\widetilde{Z}_{SM} \, .$$

We saw that the new measure we considered here does not allow for a pseudo-additive decomposition
like \ref{psadd}, and therefore it is to expect that the partition function describing the free
and mean energy cannot have the same structure $\widetilde{Z}_{SM}$ common to Sharma-Mittal
entropy unless, as can be shown \cite{Masi} applying the maximum entropy principle, one
substitutes \ref{partz} with $Z_{SE}= e_{r}^{\,\,\log \left<\frac{1}{p_{i}}\right>_{\log_{q}}}
\,.$

Summing up, the supra-extensive entropy, does no longer obey a pseudo-additive statistics, if
based on escort mean values the partition function must take an intrinsically different form than
Sharma-Mittals one, but what they have in common with R\'{e}ny's entropy is that, in general, they
do not possess the property of concavity, Lesche-stability and finite entropy production per unit
time.

\section{Conclusion}

We showed how the Sharma-Mittal and a new generalized entropy measure both unify Tsallis and
R\'{e}nyi entropies on two different paths in a way that appears natural and almost immediate when
we make use of the generalized q-logarithm an q-exponentials as in \ref{ireqt}, \ref{itqir},
\ref{eqit} and \ref{qrir}. We underlined how the relationship among all measures becomes
particularly clear using the logarithmic KN average \ref{eii} rewriting them as in
\ref{nonrelrenyi}, \ref{nonreltsallis}, \ref{nonrelnaudts} and \ref{nonrelmasi}. This path
naturally leads to the supra-extensive entropy which does not emerge from the KN means approach
alone and does not conform to a pseudo-additive law, lacks of concavity, Lesche-stability and
finite entropy production. However, because the new measure here proposed emerges so naturally as
another possible extension of R\'{e}nyi and Tsallis entropy it is therefore worth of being
mentioned. It is tempting to conclude that, while it might not have applications in a generalized
thermostatistics, it nevertheless might be of some interest in the frame of information theory,
cybernetics, control theory, etc.

Finally we obtained a way of understanding all these entropy measures in a unified picture that
can be summarized in the following table and diagram.

\newpage

\hspace{-20mm}
\begin{tabular}{|c|c|c|c|c|}
  \hline
  \textit{Entropy measure} & \textit{Explicit form} & \textit{KN-mean} form& \textit{$KN_{log}$-mean form} & \textit{$log_{q} \times exp_{q}$-form} \\
  \hline  \hline
  Supra-extensive & $\frac{\left[ 1 + \frac{(1-r)}{(1-q)} \,
  \log \sum_{i}p_{i}^{q}\right]^{\frac{1-q}{1-r}}-1}{1-q}$ &
  $\log_{q}e_{r}^{\left<\log\left(\frac{1}{p_{i}}\right)\right>_{\!exp}}$ &
  $\log_{q}e_{r}^{\log\left<\frac{1}{p_{i}}\right>_{\!\!\log_{q}}}$ &
  $\log_{q}e_{r}^{S_{R}(P,\,q)}$ \\
  \hline
  Sharma-Mittal & $\frac{1}{1-r} \left[ \left( \sum_{i} p_{i}^{\,q} \right) ^{\frac{1-r}{1-q}} - 1
  \right]$ & $\left<\log_{r}\left(\frac{1}{p_{i}}\right) \right>_{\!\!\mathrm{q-exp}}$ & $\log_{r} \left<\frac{1}{p_{i}}\right>_{\!\!\log_{q}}$ & $\log_{r} e_{q}^{S_{T}(P,\,q)}$ \\
  \hline
  Tsallis & $\frac{\sum_{i=1}^{N}p_{i}^{q}-1}{1-q}$ & $\left<\log_{q}\left(\frac{1}{p_{i}}\right) \right>_{\!\!\mathrm{lin}}$ & $\log_{q} \left<\frac{1}{p_{i}}\right>_{\!\!\log_{q}}$ & $\log_{q}e^{S_{R}(P,\,q)}$ \\
  \hline
  R\'{e}nyi  & $\frac{1}{1-q} \log \sum_{i=1}^{N} p_{i}^{q}$ & $\left<\log \left(\frac{1}{p_{i}}\right) \right>_{\!\!\mathrm{exp}}$ & $\log \left<\frac{1}{p_{i}}\right>_{\!\!\log_{q}}$ & $\log e_{q}^{S_{T}(P,\,q)}$ \\
  \hline
  BG-Shannon & $- \sum^{N}_{i=1} p_{i} \log p_{i}$ & $\left<\log \left(\frac{1}{p_{i}}\right) \right>_{\!\!\mathrm{lin}}$ & $\log \left<\frac{1}{p_{i}}\right>_{\!\!\log}$ & $\log e^{\,S_{S}(P)}$ \\
  \hline
\end{tabular}

\setlength{\unitlength}{1mm}

\begin{picture}(30,65)(15,0)

\put(0,-85){\framebox(168,130)}

\put(10,32){\large Sharma-Mittal  \normalsize}

\put(14,20){\large$\log_{r} \left<\frac{1}{p_{i}}\right>_{\!\!\log_{q}}$ \normalsize}

\put(26,16){\line(1,-1){10}}

\put(35,3){\normalsize $r \rightarrow 1$ \normalsize}

\put(41,1){\vector(1,-1){10}}

\put(43,21){\line(1,0){10}}

\put(53,21){\line(2,-1){10}}

\put(63,13.5){\normalsize $r \rightarrow q$ \normalsize}

\put(71,12){\vector(2,-1){40}}

\put(125,32){\large Supra-extensive  \normalsize}

\put(127,20){\large$\log_{q}e_{r}^{\log\left<\frac{1}{p_{i}}\right>_{\!\!\log_{q}}}$ \normalsize}

\put(140,16){\line(-1,-1){10}}

\put(125,3){\normalsize $r \rightarrow 1$ \normalsize}

\put(125,1){\vector(-1,-1){10}}

\put(122,21){\line(-1,0){12}}

\put(110,21){\line(-2,-1){10}}

\put(92,13.5){\normalsize $r \rightarrow q$ \normalsize}

\put(95,12){\vector(-2,-1){40}}

\put(46,-14){\large R\'{e}nyi \normalsize}

\put(43,-22){\large \shortstack{$\log \left<\frac{1}{p_{i}}\right>_{\!\!\log_{q}}$} \normalsize}

\put(53,-28){\line(1,-1){10}}

\put(61,-41.5){\normalsize $q \rightarrow 1$ \normalsize}

\put(68,-43){\vector(1,-1){10}}

\put(109,-14){\large Tsallis \normalsize}

\put(105,-22){\large \shortstack{$\log_{q} \left<\frac{1}{p_{i}}\right>_{\!\!\log_{q}}$}
\normalsize}

\put(114,-28){\line(-1,-1){10}}

\put(97,-41.5){\normalsize $q \rightarrow 1$ \normalsize}

\put(100,-43){\vector(-1,-1){10}}

\put(56,-60){\large Shannon (Boltzmann-Gibbs) \normalsize}

\put(75,-70){\large \shortstack{$\log \left<\frac{1}{p_{i}}\right>_{\!\!\log}$} \normalsize}

\end{picture}

\end{document}